\documentclass[12pt]{article}

\begin{document}

\title{
{\Large \bf Heisenberg-type  structures of one-dimensional 
quantum Hamiltonians  }
}

\author{
E. M. F. Curado$^{\&,a}$, M. A. Rego-Monteiro$^{\&,b}$ and H. N. Nazareno$^{\star,c}$ \\
$^{\&}$ Centro Brasileiro de Pesquisas F\'\i sicas, \\ 
Rua Xavier Sigaud 150, 22290-180 - Rio de Janeiro, RJ, Brazil \\
$^{a}$ e-mail: eme@cbpf.br ,
$^{b}$ e-mail: regomont@cbpf.br \\
$^{\star}$ International Centre of Condensed Matter Physics - ICCMP \\
Universidade de Bras\'{\i}lia, 70919-970, Bras\'{\i}lia, Brazil \\
$^{c}$ e-mail: hugo@iccmp.br
}

\maketitle

\begin{abstract}
 
\indent
 
We construct a Heisenberg-like algebra
for the one dimensional infinite square-well 
potential in quantum mechanics. The  ladder 
operators
are realized in terms of physical operators of the
system as in the harmonic oscillator algebra.
These physical operators are
obtained with the help of variables used in a 
recently developed non commutative differential 
calculus.  
This \textquotedblleft square-well algebra\textquotedblright \,  
is an example of an algebra in a large class of 
generalized Heisenberg algebras recently constructed. 
This class of algebras also contains $q$-oscillators as a
particular case.
We also discuss the physical content of this large 
class of algebras.

\end{abstract}

\vspace{1cm}

\begin{tabbing}

\=xxxxxxxxxxxxxxxxxx\= \kill

{\bf Keywords:}  Heisenberg algebra; quantum mechanics;
quantum algebras;\\ $q$-oscillators; non-linearity;  
non commutative calculus. 

 \\

{\bf PACS Numbers:} 03.65.Fd.

\end{tabbing}

\newpage

The one-dimensional quantum harmonic oscillator is an
special system in physics for several well-known reasons. 
The algebra related to it, the Heisenberg algebra, is a 
reference tool in the second quantization approach, and 
its structure, having as generators the Hamiltonian and  
the ladder operators is used in several areas of
physics. 

In the last years, there have been an intense activity
on deformed algebras \cite{defalgebras}. A deformed algebra
is a non-trivial generalization of a well-known algebra
through the introduction of one or more complex parameters,
such that, in a certain limit of the parameters the 
well-known algebra is recovered. There have been 
several attempts to generalize Heisenberg algebra and a 
particular deformation of Heisenberg algebra, known as
$q$-oscillators \cite{qoscillators}, has attracted 
considerable attention \cite{qappli,marco1, marco2}. 
Nevertheless, in all generalizations of Heisenberg algebra, 
a clear comprehension of the physical problem under 
consideration is always lacking.

Recently, it was constructed a generalization of 
the Heisenberg algebra depending on a general functional 
of one generator of the algebra, $f(J_0)$, 
\cite{algebra1,algebra2} the characteristic function
of the algebra. For linear $f$ it was shown 
that the algebra corresponds to $q$-oscillators, 
the Heisenberg algebra being obtained in the limit when 
the deformation parameter $q \rightarrow 1$. The 
representations of the algebra, when $f$ is any 
analytical function, was shown to be obtained through
the study of the stability of the fixed points of $f$ and 
of their composed functions, exhibiting an unsuspected 
link between algebraic and dynamical system formalisms.

We show here that this generalization of the Heisenberg
algebra together with a non-commutative differential 
calculus, developed to be used in space-time discrete 
networks \cite{dimakis1,dimakis2,dimakis3}, are 
appropriate to describe algebraic aspects of a simple 
quantum mechanical system: the one-dimensional infinite 
square-well potential. The generators of the 
Heisenberg-type algebra that describes the one-dimensional
square-well potential are the Hamiltonian and the
ladder operators. The ladder operators are realized in 
terms of physical operators of the system in a similar 
way to what happens in the harmonic oscillator. 

As will
be clear in what follows, the approach we have developed
here exhibits the physical content of the class of generalized
Heisenberg algebras constructed in \cite{algebra2}, i.e,
this class of algebras describes the Heisenberg-type
algebras of a class of one-dimensional quantum systems 
having energy eigenvalues ($\epsilon_{n}$) written as 
$\epsilon_{n+1} = f(\epsilon_{n})$, where 
$f(x)$ is a different function for each physical
system. This function $f(x)$ is exactly the 
characteristic function of the algebra.
 
In the final remarks we will show that 
the introduction of a Heisenberg-type algebra having
as generators the Hamiltonian of a one-dimensional
quantum system and ladder operators, 
realized in terms of physical operators of the physical 
system, opens the possibility of constructing alternative 
formalisms of second quantization with possible applications
ranging from condensed matter to quantum field theories. 
We sketch there the construction of a non-standard 
non-relativistic free quantum field theory based
on the square-well potential.

The generalization of the Heisenberg algebra recently developed in
\cite{algebra1,algebra2} can be described by the generators 
$J_{0}$, $J_{\pm}$ satisfying the relations: 
\begin{eqnarray}
J_{0} \, J_{+} &=& J_{+} \, f(J_{0}) \, , 
\label{eq:j+g} \\
J_{-} \, J_{0} &=& f(J_{0}) \, J_{-} \, , 
\label{eq:j-g} \\
\left[ J_{+},J_{-} \right] &=& J_{0}-f(J_{0}) \, ,
\label{eq:j0g}
\end{eqnarray}
where, by hypothesis $J_{-}=J_{+}^{\dagger}$, $J_{0}^{\dagger}=J_{0}$ 
and $f(J_{0})$
is a general analytic function of $J_{0}$ that we call the
characteristic function of the algebra. It can easily be shown that
the Jacobi identity of this algebra is trivially satisfied $\forall 
f$ analytic function.
The above algebraic relations 
are constructed in order that, in the representation theory, 
the $n$-th eigenvalue
of the operator $J_{0}$ is given by the $n$-th iteration,  
through the function $f$, of an initial value $\alpha_{0}$. 
The operator
\begin{equation}
C = J_{+} \, J_{-} - J_{0} = J_{-} \, J_{+} - f(J_{0}) , 
\label{eq:casimir}
\end{equation}
is a Casimir operator of the algebra. The representation theory 
of the algebra can be analyzed assuming that we have an 
irreducible representation of the algebra given by eqs. (1-3). 
Consider the state $|0\rangle$ with the lowest 
eigenvalue of  the Hermitian operator $J_{0}$, 
\begin{equation}
J_{0} \, |0\rangle = \alpha_{0} \, |0\rangle \, .
\label{vacuum}
\end{equation}
For each value of $\alpha_{0}$ and the parameters of the algebra
we have a different vacuum that for simplicity will be denoted by
$|0\rangle$. As $|0\rangle$ is the vacuum, we require
\begin{equation}
J_{-} \, |0\rangle = 0 .
\label{vacuum2}
\end{equation}
As consequence of the algebraic relations (1-3, 5, 6) we obtain for a general
functional $f$  
\begin{eqnarray}
J_{0} \, |m\rangle &=& f^{m}(\alpha_0) \, |m\rangle , \; \; \; m = 0,1,2, 
\cdots \; , 
\label{fm} \\
J_{+} \, |m\rangle &=& N_{m} \, |m+1\rangle , 
\label{nm} \\
J_{-} \, |m\rangle &=& N_{m-1} \, |m-1\rangle ,
\label{nm1}
\end{eqnarray}
where $N_{m}^2 = f^{m+1}(\alpha_0)-\alpha_0$ 
and we have used $f^0(\alpha_{0}) = \alpha_{0} $. 
Note that $f^m(\alpha_0)$
denotes the $m$-th iterate of $f$, 
\begin{equation}
\alpha_m \equiv f^m(\alpha_0) = f(\alpha_{m-1}) \, .
\label{recursion}
\end{equation}

Eqs. (7-9) define the general conditions for a 
$n$-dimensional representation of the algebra. In order to
solve it, i.e., to construct the conditions under which we have 
finite- and infinite-dimensional representations we have to specify 
the functional $f(J_0)$. The Heisenberg algebra
is the simplest particular case of algebra (1-3) and we
can see that if we choose 
$f(J_0) = J_0 + 1$ the algebra given by eqs. (1-3) becomes 
the Heisenberg algebra. In \cite{algebra2} we used  
linear and quadratic functionals, leading to multi parametric 
deformations of the Heisenberg algebra.  Also, we showed in \cite{algebra2}
that it is the 
iteration aspect of the algebra that  
allow us to find their representations through the 
analysis of the 
stability of the fixed points of the function $f$ and their
composed functions \cite{algebra1,algebra2}. 

Here, in this paper, we shall use the inverse approach 
utilized in \cite{algebra1,algebra2}, where it was 
studied the representation theory for general functions $f$. 
Now, we consider a simple one-dimensional quantum system
with a known spectrum and obtain the characteristic function
$f(x)$ of the associated Heisenberg-type algebra for
this physical problem. Moreover, we also realize the 
generators of the algebra in terms of the physical operators
of the system.
To implement this 
program we shall need the formalism of the non commutative 
differential calculus mainly studied by Dimakis et al 
\cite{dimakis1,dimakis2,dimakis3}. 

In  \cite{dimakis1} 
a formalism was developed for a one-dimensional 
spacial lattice with finite 
spacing, i.e., a discrete space. We shall summarize here an 
analogous  
formalism for a momentum-space instead of the 
position-space. The 
reason is that in many physical problems the momentum space is 
already discretized. In the 
one-dimensional infinite square-well potential for example, 
that will be analyzed below, 
the allowed values for the (adimensional) momenta are 
only the positive integers, as it is well-known. 
Thus, the non commutative differential calculus 
approach seems to be appropriate to be used in the momentum space.  
The formulae used here are analogous to the formulae used in 
\cite{dimakis1}, and the reader should see this paper for a 
more detailed exposition and explanation of the non commutative 
calculus (remembering again that their formulae 
were derived for a discrete position-space). Therefore, 
let us consider an  
one dimensional lattice in a momentum 
space where the 
momenta are allowed only to take discrete values, say $p_{0}$, 
$p_{0}+a$, $p_{0}+2a$, $p_{0}+3a$ etc, with $a>0$.  
The 
non commutative differential calculus is based on the expression 
\begin{equation}
    [p,dp] = dp \, a
    \label{eq:noncom1} \, ,
\end{equation}
implying that 
\begin{equation}
    f(p) \, dg(p) = dg(p) \, f(p+a) \, ,
    \label{eq:noncom2}
\end{equation}
for all functions $f$ and $g$. Let us introduce partial 
derivatives by 
\begin{equation}
    d \, f(p) = dp \, (\partial_{p} \, f) \, (p) = 
    (\bar{\partial}_{p} \, f) \, (p) \, dp \, ,
    \label{partial}
\end{equation}    
where the left and right discrete  derivatives are given by: 
\begin{eqnarray}
    (\partial_{p} \, f) \, (p) & = & \frac{1}{a} \, [f(p+a) - f(p)] \, ,
    \label{eq:partialleft}  \\
    (\bar{\partial}_{p} \, f) \, (p) & = & \frac{1}{a} \, [f(p) - f(p-a)] \, ,
    \label{eq:partialright}
\end{eqnarray}
and satisfy 
\begin{equation}
  (\bar{\partial}_{p} \, f) \, (p) = (\partial_{p} \, f) \, (p-a)  \, .
    \label{eq:partial2}
\end{equation}
The Leibniz rule for the right discrete derivative can be written as:
\begin{equation}
    (\partial_{p} \, fg) \, (p) = (\partial_{p}f) \, (p) g \, (p) + 
    f(p+a)(\partial_{p} g) \, (p) \, ,
    \label{eq:leibniz}
\end{equation}
with a similar formula for the left derivative \cite{dimakis1}.

Let us now introduce the momentum shift operators 
\begin{eqnarray}
     A & = & 1 + a \, \partial_{p} \, ,
    \label{eq:a} \\
    \bar{A}  & = & 1 - a \, \bar{\partial}_{p}
    \label{eq:abarra}  
\end{eqnarray}
which increases (decreases) the momentum value by $a$
\begin{eqnarray}
    (Af) \, (p) & = & f(p+a)
    \label{eq:af} \\
    (\bar{A} f) \, (p) & = & f(p-a)
    \label{eq:abarraf}  
\end{eqnarray}
and satisfies 
\begin{equation}
    A \, \bar{A} = \bar{A} A = 1 \, ,
    \label{eq:aabarra}
\end{equation}
where $1$ means the identity on the algebra of functions of $p$.  Let us 
now introduce the momentum operator \cite{dimakis1}
\begin{equation}
    (Pf) \, (p) = p \, f(p) \, ,
    \label{eq:momentum}
\end{equation}
($P^\dagger = P$), which returns the value of the variable 
of the function $f$.  Clearly, 
\begin{eqnarray}
    AP & = & (P+a)A
    \label{eq:ap}  \\
    \bar{A}P & = & (P-a) \bar{A} \, \, .
    \label{eq:abarrap}
\end{eqnarray}

Integrals can also be defined in this formalism but 
it is rather a technical point and the interested 
reader can find in \cite{dimakis1} a detailed 
explanation on the subject. Here we will 
only use the definition of a definite integral of a function 
$f$ from $p_{d}$ to $p_{u}$ ( $p_{u}$ being equal to 
$p_{d} + M  a$, 
where $M$ is a positive integer) as 
\begin{equation}
    \int_{p_{d}}^{p_{u}} dp \, f(p) = a \, 
    \sum_{k=0}^{M} \, f(p_{d} + k \, a) \, .
    \label{eq:integral}
\end{equation} 
Using eq. (\ref{eq:integral}), 
an inner product of two (complex) functions 
$f$ and $g$ can be defined as 
\begin{equation}
    \langle f \, , \, g \rangle = \int_{p_{d}}^{p_{u}} dp \, f(p)^{*} \, g(p) \, , 
    \label{eq:inner}
\end{equation}
where $^{*}$ indicates the complex conjugation of the function $f$.  Clearly, 
the norm $\langle f \, , \, f \rangle \geq 0$ is
zero only when $f$ is identically null.  The set of equivalent classes  
of normalizable functions $f$ ($\langle f \, , \, f \rangle $ is finite) 
is a Hilbert space and it can be shown that   
the operators $A$ and $\bar{A}$ are well defined in 
this space \cite{dimakis1}. We have
\begin{equation}
    \langle f, A g \rangle = \langle \bar{A} f, g \rangle \, ,
    \label{eq:inner2}
\end{equation}
where  
\begin{equation}
    \bar{A} = A^\dagger \, \, ,
    \label{eq:adjoint}
\end{equation}
being $A^\dagger$ the adjoint operator of $A$.  Eqs. (\ref{eq:aabarra}) and 
(\ref{eq:adjoint}) show that $A$ is a unitary operator. 
It is also possible to define a position operator $X$ given as
$X = (\partial_{p} + \bar{\partial}_{p})/2 i$ \cite{dimakis1}. 
With this very short 
adapted review of the non commutative differential calculus 
we can go further and, 
together with the generalization of the Heisenberg algebra, analyze the  
physical example of the 
quantum mechanical infinite one dimensional square-well potential. 

Thus, let us assume a one dimensional system with 
zero potential between zero 
and $L$ and infinite elsewhere. As it is well-known, the spectrum of the 
Hamiltonian ($H= c P^{\, 2}$, $c=1/2m$, $\hbar = 1$) 
with the above boundary conditions is proportional 
to $n^2$, where 
$n = 1, 2, 3, \ldots$.  The momentum is quantized and 
proportional to $n$.  Therefore, we can 
see the momentum space as an one dimensional periodic lattice 
with constant spacing $a= \pi/L$, clearly 
a candidate to apply the non commutative differential 
calculus reviewed before. We then take the momentum operator
in the Hamiltonian $H= c P^{\, 2}$, with the above boundary conditions,
as defined in eq. (\ref{eq:momentum}). 

The Hamiltonian's eigenvalue associated with the (n+1)-th level is 
proportional to $(n+1)^2$ and we can write
\begin{equation}
    e_{n+1} = b (n+1)^2 = ( \sqrt{e_{n}} + \sqrt{b} )^2 \, ,
    \label{eq:recorrencia}
\end{equation}
where $e_{n}$ is the eigenvalue of the Hamiltonian associated with 
the $n$-th level and $b= \pi^2/2mL^2$.  As $J_{0}$ is related to the 
Hamiltonian \cite{algebra1,algebra2} and their eigenvalues
satisfy the iterations given 
by a function $f$ in eqs. (\ref{eq:j+g} - \ref{eq:j0g}), we see that 
if we choose this function as
\begin{equation}
    f(x) = ( \sqrt{x} + \sqrt{b} )^2 \, ,  
    \label{eq:deff}
\end{equation}
the $J_0$ in eqs. (\ref{eq:j+g}-\ref{eq:j0g}) has eigenvalues
equal to the energy eigenvalues of the square-well potential. Eqs. 
(\ref{eq:j+g}-\ref{eq:j0g}) can then be rewritten 
for this case as 
\begin{eqnarray}
\left[ J_{0},J_{+} \right] &=& 2 \sqrt{b} \, J_{+} \, \sqrt{J_{0}} + b \, J_{+} \, , 
\label{eq:j+sw} \\
\left[ J_{0},J_{-} \right] &=& -2 \sqrt{b} \, \sqrt{J_{0}} \, J_{-}  - b \, J_{-} \, , 
\label{eq:j-sw} \\
\left[ J_{+},J_{-} \right] &=& -2 \sqrt{b} \, \sqrt{J_{0}} - b \, .
\label{eq:j0sw}
\end{eqnarray}
The square root of the generator $J_0$ is well defined since
this is a Hermitian and positive definite operator.

We then have an algebra eqs. (\ref{eq:j+sw}-\ref{eq:j0sw}) where,
by construction, the eigenvalues of $J_0$ , $e_n$, are the energy
eigenvalues of the quantum mechanical one dimensional infinite
square-well potential and $J_{\pm}$ act as ladder operators. In order
to have a complete description, similar to the case of the one-dimensional
harmonic oscillator, we must realize the operators $J_{(\pm \, ,\, 0)}$ in
terms of physical operators. We propose for this problem the
following realization: 
\begin{eqnarray}
    J_{+} & = &  (c P^2 -b)^{1/2}  \,\bar{A} 
    \label{eq:j+}  \\
     J_{-} & = & A \, (c P^2 -b)^{1/2}
    \label{eq:j-}  \\
    J_{0} & = & c \, P^2  \, .
    \label{eq:j0}
\end{eqnarray}
Clearly, $J_{0}$ is the Hamiltonian and can be written, analogously 
to the harmonic oscillator case, as an ordered product of 
ladder operators 
\begin{equation}
     J_{+} \, J_{-} = J_{0} - b \, ,
    \label{eq:joH}
\end{equation}
as according to eq. (\ref{eq:aabarra}), 
$A \, \bar{A} = 1$.  Comparing eqs. (\ref{eq:casimir}) and (\ref{eq:joH})
we see that $b$ is the Casimir of the representation for the
square-well potential.
Using eqs. (24-25) it is straightforward to 
check that these operators indeed 
satisfy the commutation relations given by eqs. (\ref{eq:j+sw}-\ref{eq:j0sw}).
We stress that, the operators $P$ and $X$
are the momentum and position operators in the momentum space
for the one-dimensional infinite square-well potential. 
Moreover, as will be seen below it is possible 
to write the operators $P$ and 
$X$ in terms of the ladder operators
$J_\pm$ and the operator $J_0$.

Fock space representations of the algebra generated by $J_{0}$ and $J_{\pm}$,  
eqs. (\ref{eq:j+sw}-\ref{eq:j0sw}), are 
obtained considering eigenstates of $J_{0}$, with fixed values 
of the momentum. Let us call $| n \rangle$ the 
eigenstate of $J_{0}$ 
whose momentum is associated with the quantum number $n$, 
$n= 1, 2, 3, \ldots$.  The eigenvalue $\alpha_{n}$ that appears in 
eqs. (\ref{vacuum}-\ref{recursion}) can be put as  
$\alpha_{n} = b \, (n+1)^2$ 
and the eqs. (\ref{fm}-\ref{nm1}) can 
be rewritten, after a trivial rename of the states $|n\rangle$ 
such that the lowest energy state corresponds to $|1\rangle$, 
as 
\begin{eqnarray}
    J_{0} \, |n\rangle &=& b \, n^2 \, |n\rangle , \; \; \; n = 1,2, 
\cdots \; ,
    \label{eq:fmsw}  \\
   J_{+} \, |n\rangle &=& \sqrt{b(n+1)^2 -b} \, \, |n+1\rangle \, , 
    \label{eq:nmsw}  \\
   J_{-} \, |n\rangle &=& \sqrt{b n^2 -b} \, \, |n-1\rangle \, ,
    \label{eq:nm1sw}  \\
    P |n\rangle &=& a \, n \, |n \rangle , \, \, \bar{A} |n\rangle =  
|n+1 \rangle \,  ,
\end{eqnarray}
where $N_{n}^2 = b \, (n+1)^2 - b$. Note that, $J_- |1\rangle = 0$
as it happens in the standard notation of the square-well 
potential since the lowest energy state is represented by the
state $|1\rangle$.
    
Hence, we see that an algebraic formalism similar to 
the harmonic oscillator algebra was constructed for another 
physical problem: the one dimensional infinite square-well 
potential in quantum mechanics. The main point here is that 
the Hamiltonian itself is one of the generators of the algebra, 
together with the ladder operators. In other physical realizations 
of the ladder operators \cite{morse}, the Hamiltonian is not, 
in general, one of the generators of the algebra. 

Generally speaking, suppose we have
an arbitrary one-dimensional quantum system such that
two successive energy eigenvalues, $\epsilon_n$, can be 
related as 
\begin{equation}
\epsilon_{n+1} = f(\epsilon_{n}) \, \, ,
\label{eq:recursion2}
\end{equation} 
where 
$f(x)$ is a different function for each physical
system. If we assume that the generator $J_0$ of 
the class of Heisenberg-type algebras in
eqs. (\ref{eq:j+g}-\ref{eq:j0g}) is the Hamiltonian
operator of this one-dimensional quantum system,
eq. (\ref{recursion}) tell us that the algebra
in eqs. (\ref{eq:j+g}-\ref{eq:j0g}) with $f$ appearing
in eq. (\ref{eq:recursion2}) describes the algebraic
structure of this quantum system. Moreover, by 
eqs. (\ref{nm}-\ref{nm1}) we see that $J_+$ and $J_-$ 
are the ladder operators of this quantum system. 
In summary, the Heisenberg-type algebras  
\cite{algebra2} given in eqs. (\ref{eq:j+g}-\ref{eq:j0g})
describe the algebraic structure of one-dimensional
quantum systems having successive eigenvalues
related by eq. (\ref{eq:recursion2}) where the
characteristic function of the algebra is the function
$f(x)$ appearing in eq. (\ref{eq:recursion2}).

Once it is understood the Heisenberg-type structure
of a one-dimensional quantum system, the next step,
as was performed here for the square-well potential,
is to realize the ladder operators of the algebra
in terms of the physical operators of the system, such 
that the algebra 
is still satisfied and being the product $J_+ \, J_-$ 
proportional to the Hamiltonian of the one-dimensional
quantum system under consideration.
This program could supply an alternative approach
to quantum field theory as indicated in what
follows.

Using the momentum operator $P$ defined
on a lattice, eq. (\ref{eq:momentum}), and the associated lattice 
derivatives we can define two type coordinate operators as
\begin{eqnarray}
    X & = & \frac{1}{2i} (\bar{\partial}_p + \partial_p) \, \, ,
    \label{eq:cord1}  \\
     Q & = &  \bar{\partial}_{p} - \partial_{p} \, \, ,
    \label{eq:cord2}  
\end{eqnarray}
where $\partial_p$ and $\bar{\partial}_p$ are the left and right 
discrete derivatives defined in eqs. (\ref{eq:partialleft}, 
\ref{eq:partialright}). Of course, in the continuous limit
($a\rightarrow 0$)
the operator $Q$ is identically null since $\partial_p$ and 
$\bar{\partial}_p$ represent, in this limit, the same derivative.
It can be checked that the operators $P$, $X$ and $Q$
generate an algebra on the momentum lattice that reduces
to the standard Heisenberg algebra when $a\rightarrow 0$.
With the help of eqs. (\ref{eq:a}-\ref{eq:abarra} and 
\ref{eq:j+}-\ref{eq:j-}) 
we can rewrite $X$ and $Q$ in terms of the ladder
operators of the $a$-deformed Heisenberg algebra as
\begin{eqnarray}
X &=&  \frac{i}{2a} \left( S^{-1} A^{\dagger} - A S^{-1} \right)  ,
\label{eq:cord3} \\
Q &=&  \frac{1}{a} \left( -2 + S^{-1} A^{\dagger} + A \, S^{-1} \right) , 
\label{eq:cord4} 
\end{eqnarray}
where $S=(c P^2 -b)^{1/2}$. Using an independent copy
of the operators $Q$ and $X$ for each point of
a three-dimensional lattice we can define two fields
and two momentum fields that can be used to construct
a free quantum field theory Hamiltonian. This Hamiltonian
can be written as 
\begin{eqnarray}
H &=& \sum_{\vec{k}} 
J_{+}(\vec{k}) J_{-}(\vec{k}) =
\sum_{\vec{k}} S_{\vec{k}}^2  \nonumber \\
&=& 
\sum_{\vec{k}} (c P_{\vec{k}}^2 -b(\vec{k}))  \,\,  ,
\label{eq:resulthamilt}
\end{eqnarray}
where $P_{\vec{k}}$, for each $\vec{k}$, is the momentum operator for a
particle with mass $m$ in a square-well potential. This is a
non-relativistic free quantum field theory and the details of 
construction of a relativistic free quantum field theory
can be found in \cite{qft}.

The eigenvectors of $H$ form a complete set and span the Hilbert
space of this system. The eigenvectors are
\begin{equation} 
|1 \rangle, \,\, J_{+}(\vec{k}) |1 \rangle, \,\,
J_{+}(\vec{k}) J_{+}(\vec{k}') |1 \rangle \,\,
\mbox{for} \,\, \vec{k}\not= \vec{k}', \,\,
(J_{+}(\vec{k}))^2 |1 \rangle, \,\, \cdots
\label{eq:hilbert} 
\end{equation}
This Hilbert space has a different interpretation with
respect to the standard spin-$0$ quantum field theory based
on the harmonic oscillator. While in the standard
quantum field theory the creation operator creates one
particle of mass $m$ each time it is applied to the vacuum, 
in this non-relativistic 
quantum field theory one reads from eq. 
(\ref{eq:resulthamilt}) that the creation operator in this
case creates excited states of a particle in a box. This could
provide an alternative quantum field theory phenomenological 
approach to hadronic interactions.

\vspace{0.7 cm}
\noindent
{\bf Acknowledgments:} The authors thank S. Sciuto, C. Tsallis and
L. Castellani for useful comments. E. M. F. C. and 
M. A. R. M., thank PRONEX for partial support. E.M.F.C. and H.N. Nazareno 
thank CNPQ for a grant.

\end{document}